\documentclass[journal]{IEEEtran}

\ifCLASSINFOpdf
\else
   \usepackage[dvips]{graphicx}
\fi

\usepackage{graphicx}
\usepackage{booktabs}
\usepackage{multirow}
\usepackage{url}
\usepackage{cite}
\usepackage{amsmath}

\begin{document}

\title{Noise-Robust Hearing Aid Voice Control}

\author{Iv\'an L\'opez-Espejo, Eros Rosell\'o, Amin Edraki, Naomi Harte, and Jesper Jensen\thanks{This work has been funded by the Spanish Ministry of Science and Innovation under the ``Ram\'on y Cajal'' programme (RYC2022-036755-I).}
\thanks{I. L\'opez-Espejo and E. Rosell\'o are with the Dept. of Signal Theory, Telematics and Communications, UGR, Spain (e-mails: \{iloes,erosrosello\}@ugr.es).}
\thanks{A. Edraki is with the Dept. of Electrical and Computer Engineering, QU, Canada (e-mail: a.edraki@queensu.ca).}
\thanks{N. Harte is with the School of Eng., TCD, Ireland (e-mail: nharte@tcd.ie).}
\thanks{J. Jensen is with Oticon A/S, Denmark, as well as with the Dept. of Electronic Systems, AAU, Denmark (e-mail: jesj@demant.com).}}

\markboth{Accepted by IEEE Signal Processing Letters}
{Shell \MakeLowercase{\textit{et al.}}: Bare Demo of IEEEtran.cls for IEEE Journals}
\maketitle

\begin{abstract}
Advancing the design of robust hearing aid (HA) voice control is crucial to increase the HA use rate among hard of hearing people as well as to improve HA users' experience. In this work, we contribute towards this goal by, first, presenting a novel HA speech dataset consisting of noisy own voice captured by 2 behind-the-ear (BTE) and 1 in-ear-canal (IEC) microphones. Second, we provide baseline HA voice control results from the evaluation of light, state-of-the-art keyword spotting models utilizing different combinations of HA microphone signals. Experimental results show the benefits of exploiting bandwidth-limited bone-conducted speech (BCS) from the IEC microphone to achieve noise-robust HA voice control. Furthermore, results also demonstrate that voice control performance can be boosted by assisting BCS by the broader-bandwidth BTE microphone signals. Aiming at setting a baseline upon which the scientific community can continue to progress, the HA noisy speech dataset has been made publicly available.
\end{abstract}


\IEEEpeerreviewmaketitle

\section{Introduction}
\label{sec:intro}

\IEEEPARstart{H}{earing} loss is a ubiquitous problem that, according to the World Health Organization, will affect 1 in 4 people by 2050 \cite{WHO}. To mitigate the impacts of hearing loss, hearing aids (HAs) have the potential to help hard of hearing people improve their daily life engagement, confidence, and overall life quality \cite{Ruth24}. Despite this, many hard of hearing people who could benefit from a HA do not wear one due to stigma and/or aesthetic reasons \cite{Forbes,Aesthetics}. Hence, advancing the design of invisible, in-ear-canal (IEC) HAs is key to increase their rate of use among people who need them.

To manipulate invisible HAs, voice control implemented via keyword spotting (KWS) \cite{KWSOverview} can be an extremely useful technology. In addition, voice control can also be a highly valuable tool to handle visible, behind-the-ear (BTE) HAs \cite{Lopez19,Lopez20}, especially for elderly people who have a hard time manipulating the device's small buttons \cite{KWSOverview}.

Motivated by the above facts, this paper dives into essentially uncharted waters aiming at setting a baseline for noise-robust HA voice control upon which the scientific community can continue to advance. This is achieved through the following two contributions.
\begin{enumerate}
    \item The creation of a one-of-a-kind speech dataset ---based on the Google Speech Commands Dataset (GSCD) \cite{Warden18}--- containing (simulated) noisy own voice as if it were captured by a HA with 2 (front and rear) BTE microphones and 1 IEC microphone.
    \item The training and evaluation of light, state-of-the-art KWS models \cite{Kim21} for HA voice control that leverage different combinations of microphone signals from the proposed HA noisy speech dataset.
\end{enumerate}

In relation to the proposed dataset\footnote{Note that the underlying hearing aid-related data originate from \cite{Edraki24}.}, which has been made publicly available\footnote{\url{https://zenodo.org/records/14140800}}, it is interesting to note that the IEC microphone is placed inside an ear occluded by an acoustic seal. This occlusion leads to bone-conduction vibrations originating from own voice to resonate inside the ear canal (\emph{bone-conducted speech}, BCS). These vibrations are then captured by the IEC microphone with a relatively high signal-to-noise ratio (SNR), since BCS is not greatly contaminated by ambient noise due to the passive attenuation of the acoustic seal \cite{Rachel17}. Nevertheless, due to transmission through human tissues, BCS suffers from a limited bandwidth (i.e., up to 2-2.5 kHz \cite{Edraki24}) as well as from low-frequency distortion.

Given the noise-robustness properties inherent to BCS, assisting regular air-conducted speech by BCS is proven to boost the performance of several speech technology applications like speech enhancement \cite{Garudadri18}, automatic speech recognition \cite{Susanto23}, and human sound classification \cite{Liang22}. In this work, we, for the first time, quantify the benefits of exploiting BCS for noise-robust HA voice control. Specifically, we demonstrate how a light KWS model ($\sim$50k parameters) \cite{Kim21} (which might be embedded into a HA) is able to provide a strong performance when employing the IEC microphone instead of the BTE ones, especially at highly-adverse SNRs. Furthermore, we also show that voice control performance can be significantly enhanced by complementing noise-robust BCS with the broader-bandwidth BTE microphone signals.

The rest of this paper is organized as follows. Section \ref{sec:database} explains how the HA noisy speech dataset is created. The KWS system used for HA voice control is described in Section \ref{sec:framework}. KWS results on the HA speech dataset are discussed in Section \ref{sec:results}. Finally, Section \ref{sec:conclusion} concludes this work.

\section{Hearing Aid Noisy Speech Dataset}
\label{sec:database}

This section describes the proposed HA noisy speech dataset. First, the setup to measure HA-related transfer functions and how these are used to generate HA noisy own voice from mouth reference microphone (MRM) clean speech is presented in Subsecs. \ref{ssec:ha_setup} and \ref{ssec:generation}, respectively. Then, Subsec. \ref{ssec:structure} explains how the dataset is created and structured.

While there are some similarities between HA datasets reported in \cite{Lopez19,Lopez20} and the one proposed here, this new corpus is the only one to include both noisy speech and an IEC microphone.

\subsection{Hearing Aid Transfer Function Measurement Setup}
\label{ssec:ha_setup}

\begin{figure}
\centerline{\includegraphics[width=0.82\columnwidth]{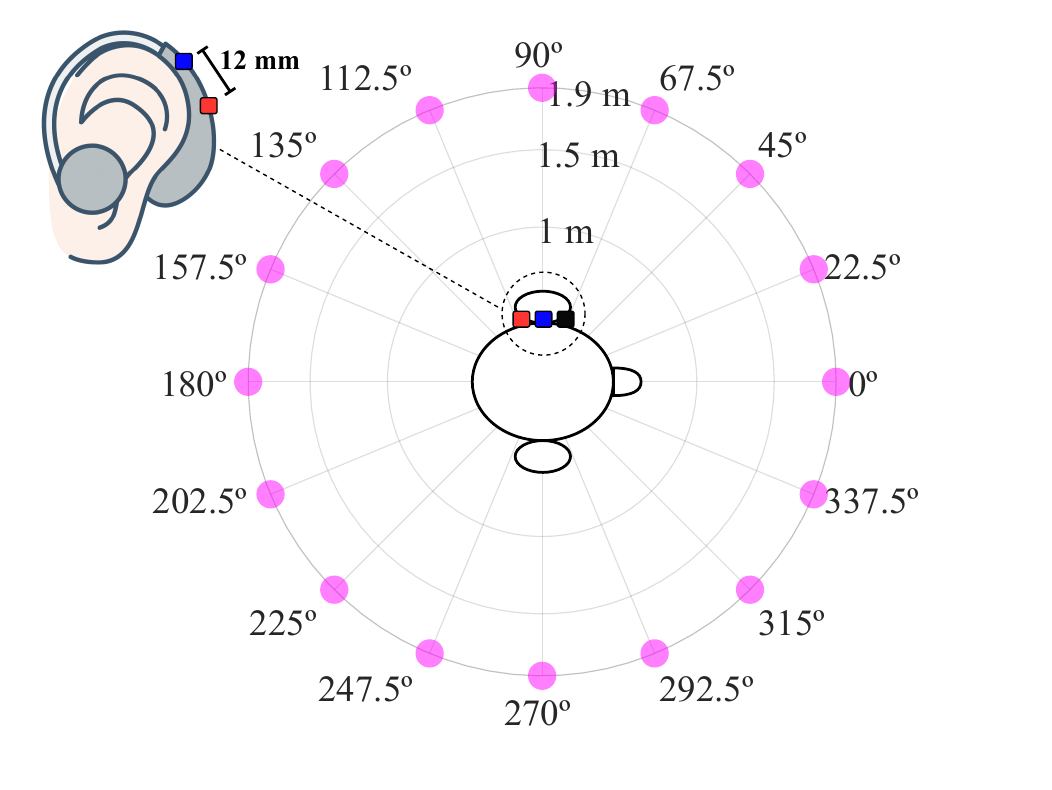}}
\caption{Experimental setup for transfer function measurement depicting a subject wearing a left ear HA that embeds IEC, front and rear microphones (black, blue and red dots, respectively). The subject is surrounded by 16 equidistantly-spaced loudspeakers at eye-height (magenta dots).}
\label{fig:setup}
\end{figure}


Own-voice transfer functions (OVTFs) and head-related transfer functions (HRTFs) were estimated from the experimental setup depicted in Fig. \ref{fig:setup}. One at a time, a total of 22 subjects sat in a soundproof sound studio while wearing HA prototypes\footnote{For experimental purposes, and assuming symmetry, we just consider the left ear HA, as shown in Fig. \ref{fig:setup}.} by Demant A/S that embed 2 (front and rear) BTE microphones and 1 IEC microphone. Subjects were surrounded by 16 equidistantly-spaced loudspeakers laid out in a circular array at eye-height.

With this experimental setup, subjects read out random Wikipedia passages aloud so that an own-voice dataset could be collected \cite{Edraki24}. Own voice was captured by all the HA microphones as well as by an MRM placed in front of each subject's mouth. Then, subject-wise OVTFs between the MRM and each of the HA microphones were estimated using a least mean squares approach.

Additionally, sweep signals consisting of pure sinusoidal signals with frequency changing from near DC to slightly above 20 kHz were emitted by each loudspeaker around the subjects and recorded by the HA microphones. These signals were used to estimate subject-wise HRTFs between each loudspeaker and each of the HA microphones \cite{Farina00}.

\subsection{Noisy Own-voice Generation Model}
\label{ssec:generation}

Given a clean speech signal $x(n)$ captured at a HA user's mouth, our aim is to simulate noisy versions of $x(n)$ at the IEC, front and rear HA microphones. We denote these noisy own-voice signals as $y_{m}(n)$, where $m\in\{\mathsf{i},\mathsf{f},\mathsf{r}\}$, and $\mathsf{i}$, $\mathsf{f}$ and $\mathsf{r}$ refer to the IEC, front and rear microphones, respectively.

Assuming a linear time-invariant system, the clean own-voice signal $x(n)$ at microphone $m$ can be calculated, for all $m$, as $x_{m}(n)=h^{m}(n) \ast x(n)$, where $h^{m}(n)$ is a user-specific OVTF between the mouth and microphone $m$, and $\ast$ denotes convolution. In addition, clean own voice is contaminated by additive noise coming from \emph{any} $L$ (which is an integer between 1 and 16) out of the 16 point noise sources represented by the loudspeakers of Fig. \ref{fig:setup}. Let $\{\nu_\ell(n)\}_{\ell=1}^{L}$ and $\{g_\ell^{m}(n)\}_{\ell=1}^L$ be the set of $L$ noise signals at their sources and user-specific HRTFs between such sources and microphone $m$, respectively. Then, the noisy own-voice signal at microphone $m$, $y_{m}(n)$, is approximated, for all $m$, as
\begin{equation}
    y_{m}(n)=x_{m}(n)+\alpha\displaystyle\sum_{\ell=1}^{L}g_\ell^{m}(n) \ast \nu_\ell(n),
    \label{eq:model}
\end{equation}
where $\alpha$ is a noise scaling factor. Note that we do not account for the Lombard effect despite it is known to have an impact on speech recognition performance \cite{Marxer18}.

\subsection{Dataset Creation and Structure}
\label{ssec:structure}

\begin{table*}
\caption{Description of the types of noise used to generate the HA noisy speech dataset. SSN stands for speech-shaped noise. Babble, SSN, and interfering speaker noises were built from the TIMIT dataset \cite{timit} such that TIMIT speakers did not overlap across the training, validation and test sets of the proposed HA dataset}
\label{tab:noises}
  \centering
  \resizebox{\linewidth}{!}{
  \begin{tabular}{c|c|l}
    \toprule[1pt]\midrule[0.3pt]
    \textbf{Noise type} & \textbf{Set} & \textbf{Description} \\ \midrule
    Babble & Train, Val \& Test & 5 female and 5 male speakers randomly located (utterance-wise) around the HA user \\
    Music & Train, Val \& Test & Stereo songs played through two adjacent, randomly picked (utterance-wise) loudspeakers around the HA user \\
    SSN & Train, Val \& Test & Surround SSN played through all the loudspeakers around the HA user \\ \hline
    Interfering speaker & Test & A female or male (same probability) speaker randomly located (utterance-wise) around the HA user \\
    TV & Test & A TV program (America's Got Talent) played through the loudspeaker at 0º \\
    \midrule[0.3pt]\bottomrule[1pt]
  \end{tabular}}
\end{table*}

\begin{figure*}
\centerline{\includegraphics[width=\textwidth]{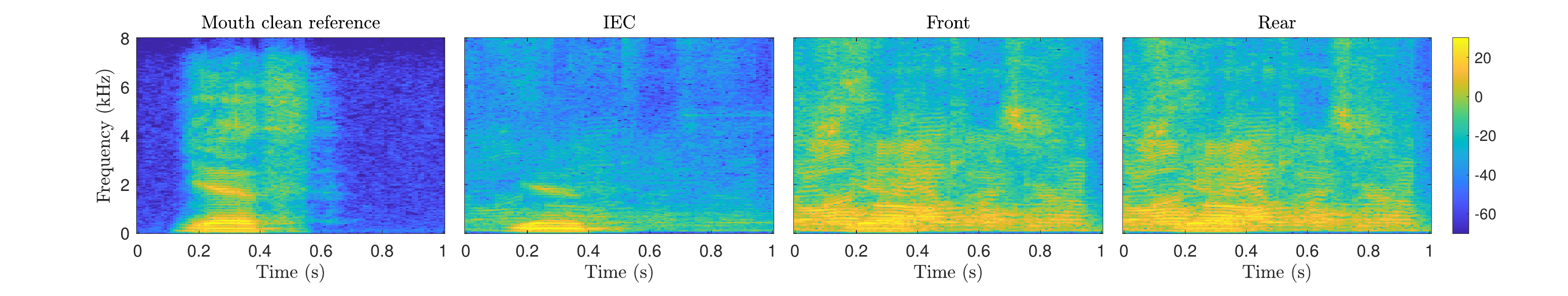}}
\caption{Magnitude spectrograms (in dB) from an amplitude-normalized test utterance comprising the word ``yes'' which is contaminated by babble noise at a 0 dB SNR at the HA's front microphone. From left to right: mouth clean reference, and IEC, front and rear microphones.}
\label{fig:specs}
\end{figure*}


Mouth reference microphone signals $x(n)$ were obtained from the GSCD \emph{v2} \cite{Warden18}, which is composed of more than 100k 1-second long utterances comprising one word each. Similar to \cite{Lopez21b}, we first extracted clean training, validation and test sets out of the GSCD by picking all the signals in each set with an \emph{a posteriori} SNR above 40 dB. Twelve classes (which are approximately balanced across the training, validation and test sets) were defined: the 10 standard keywords of the GSCD (``yes'', ``no'', ``up'', ``down'', ``left'', ``right'', ``on'', ``off'', ``stop'' and ``go''), the filler class (comprised of the other 25 GSCD words), and ambient noise (i.e., no own voice is present).

While we were provided with transfer functions from 22 subjects (see Subsec. \ref{ssec:ha_setup}), only 5 of the subjects showed good ear occlusion (which was determined by visual inspection of HRTFs) and were kept for noisy own-voice generation. From those 5, 3 were allocated to the training set, 1 more, to the validation set, and the final one, to the test set. In turn, each GSCD training speaker was randomly assigned 1 out of the 3 training HA subjects.

Following the generation model of Subsec. \ref{ssec:generation}, the clean GSCD utterances were convolved with the corresponding OVTFs to create clean own-voice signals at the IEC, front and rear microphones. To improve the generalization ability of the neural models, \emph{training} OVTFs were perturbed prior to convolution on an utterance- and microphone-basis as in \cite{Lopez20}:
\begin{equation}
    \tilde{h}^{m}(n)=(1+\gamma_n)h^{m}(n)+\delta_n,
    \label{eq:augmentation}
\end{equation}
where $m\in\{\mathsf{i},\mathsf{f},\mathsf{r}\}$, $\gamma_n\sim\mathcal{N}\left(\mu=0,\sigma=0.1\right)$ and $\delta_n\sim\mathcal{N}\left(\mu=0,\sigma=10^{-5}\right)$.

Inspired by the structure of the classical AURORA-2 database \cite{Pearce00}, the training, validation and test sets were split into 3, 3 and 5 even partitions, respectively. Each of these partitions was distorted with a different noise type (see Table \ref{tab:noises}) by considering Eq. (\ref{eq:model}), where \emph{training} HRTFs were also perturbed as in Eq. (\ref{eq:augmentation}). Copies of these partitions were generated at different front-microphone SNR levels by adjusting $\alpha$ in Eq. (\ref{eq:model}) according to the active speech level of $x_{\mathsf{f}}(n)$ in line with the ITU P.56 recommendation \cite{P56}. Note that $\alpha$ is constant across $y_{\mathsf{i}}(n)$, $y_{\mathsf{f}}(n)$ and $y_{\mathsf{r}}(n)$. Training and validation SNRs are $\{-15, -5, 5, 15, 25\}$ dB, whereas test SNR values are $\{-18, -9, 0, 9, 18\}$ dB. These SNR ranges were roughly designed by taking into account expected own-voice and acoustic noise sound pressure levels as well as an approximate mouth-to-ear loss \cite{Comsol}. Finally, note that \emph{every} utterance was contaminated with a distinct noise realization.

Roughly, the training, validation and test sets contain 18.8 hrs, 2.1 hrs and 2.4 hrs, respectively, of multi-channel audio.

For illustrative purposes, Fig. \ref{fig:specs} displays magnitude spectrograms from a test utterance containing the word ``yes'' which is distorted by babble noise at a front-microphone SNR level of 0 dB. As can be seen, the IEC microphone signal comprising BCS is significantly less affected by noise but shows a more limited bandwidth than the front and rear microphone signals.

\section{Keyword Spotting System for Voice Control}
\label{sec:framework}

Low-complexity KWS systems may be deployed in HAs to control them via voice \cite{KWSOverview,Lopez19,Lopez20}. In the next two subsections, the KWS model used in this work is introduced along with its training details.

\subsection{Keyword Spotting Model}
\label{ssec:model}

We use a light KWS model called \texttt{BC-ResNet} \cite{Kim21} that provides state-of-the-art results on the GSCD benchmark \cite{KWSOverview}. \texttt{BC-ResNet} is based on broadcasted residual learning which exploits the benefits of 1D temporal and 2D convolutions while assuring low computational complexity. In this paper, we test 6 variants of \texttt{BC-ResNet} which are denoted as \texttt{BC-ResNet-$\tau$}, where $\tau\in\{1, 1.5, 2, 3, 6, 8\}$ refers to the factor by which the base number of model channels is multiplied. The larger $\tau$ is, the more parameters the model has.

\texttt{BC-ResNet} is fed with 40-dimensional log-Mel feature vectors computed by means of a 30 ms window with a 10 ms shift across 1-second long utterances. In case of evaluating HA microphone combinations, microphone-wise log-Mel feature matrices are stacked across the channel dimension before being input to the KWS model.

\subsection{Model Training Details}
\label{ssec:training}

As in \cite{Kim21}, \texttt{BC-ResNet} is trained for a total of 200 epochs to recognize the 12 classes stated in Subsec. \ref{ssec:structure}, i.e., 10 different keywords, filler and ambient noise. As an optimizer, stochastic gradient descent with a weight decay of $10^{-3}$ and a momentum factor of 0.9 is employed. During the first 5 epochs, the learning rate linearly increases from 0 to 0.1 as warm-up and then decays to 0 with cosine annealing \cite{Hutter17}. The size of the mini-batch is 100 training utterances.

Finally, note that, different from \cite{Kim21}, we consider neither SpecAugment \cite{SpecAugment} nor any other data augmentation method.

\begin{figure}
	\centerline{\includegraphics[width=\columnwidth]{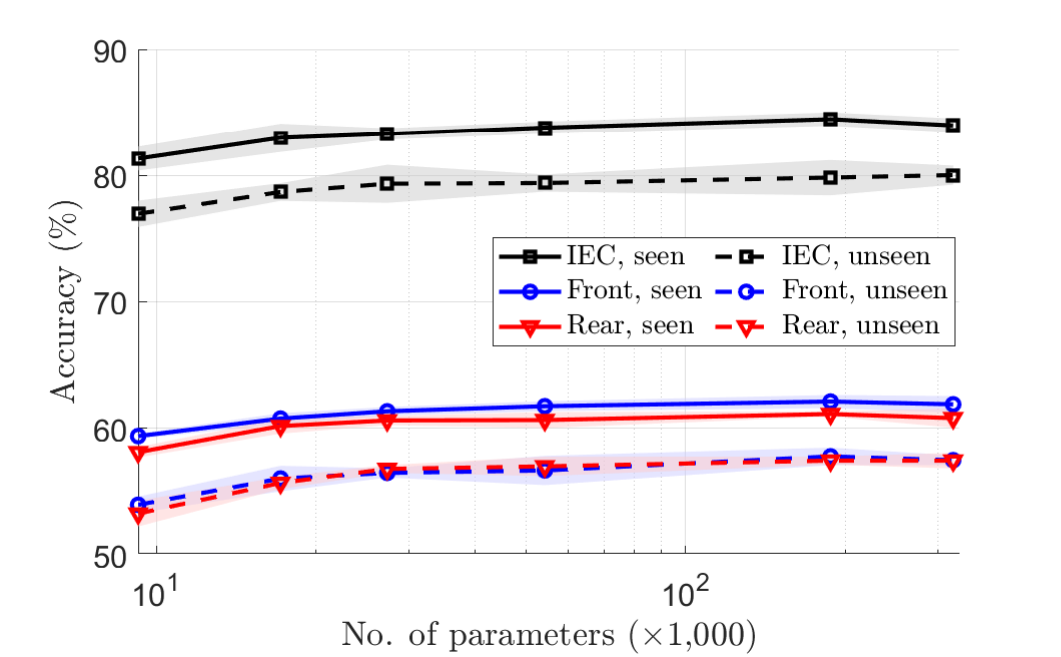}}
	\caption{Average KWS accuracy (\%) as a function of the number of model parameters when testing the use of the different HA microphones. Shaded areas represent 95\% confidence intervals.}
	\label{fig:single}
\end{figure}

\begin{figure}
	\centerline{\includegraphics[width=\columnwidth]{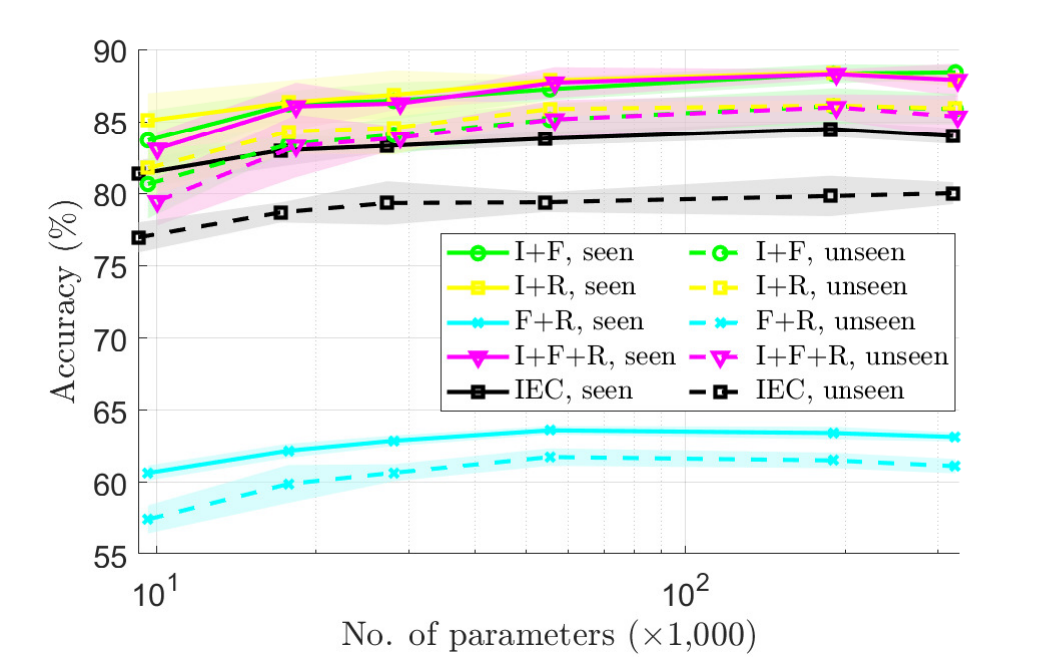}}
	\caption{Average KWS accuracy (\%) as a function of the number of model parameters when testing HA microphone combinations. IEC curves are shown as a reference. Shaded areas represent 95\% confidence intervals.}
	\label{fig:combo}
\end{figure}

\begin{table*}
	\caption{Average KWS accuracy (\%) with 95\% confidence intervals and RTF results obtained by \texttt{BC-ResNet-3} employing different HA microphones. KWS results are broken down by SNR and seen/unseen noises. Best results are marked in boldface}
	\label{tab:results}
	\centering
	\begin{tabular}{l|c|ccccc|ccccc}
		\toprule[1pt]\midrule[0.3pt]
		\multirow{2}{*}{\textbf{Mic(s)}} & \multirow{2}{*}{\textbf{RTF}} & \multicolumn{5}{c|}{SNR (dB) -- \textit{Seen noises}} & \multicolumn{5}{c}{SNR (dB) -- \textit{Unseen noises}} \\
		& & -18 & -9 & 0 & 9 & 18 & -18 & -9 & 0 & 9 & 18 \\ \midrule
		IEC & \multirow{3}{*}{0.084} & 50.7$\pm$2.6 & 86.0$\pm$0.6 & 92.9$\pm$1.0 & 94.5$\pm$0.4 & \textbf{95.1$\pm$0.4} & 37.2$\pm$2.6 & 81.3$\pm$0.9 & 90.6$\pm$0.3 & 93.6$\pm$0.8 & 94.2$\pm$0.4 \\
		Front & & 13.4$\pm$0.5 & 39.7$\pm$1.8 & 73.5$\pm$1.0 & 89.1$\pm$0.5 & 93.0$\pm$0.9 & 13.5$\pm$1.2 & 34.3$\pm$2.4 & 65.4$\pm$1.3 & 82.2$\pm$1.4 & 88.0$\pm$1.5 \\
		Rear & & 13.1$\pm$0.9 & 36.9$\pm$1.1 & 72.0$\pm$1.1 & 87.7$\pm$1.1 & 93.5$\pm$0.9 & 13.1$\pm$0.9 & 34.5$\pm$1.4 & 64.7$\pm$2.1 & 83.1$\pm$1.3 & 89.7$\pm$1.2 \\
		\midrule[0.3pt]
		I+F & \multirow{3}{*}{0.086} & 66.7$\pm$1.9 & 87.3$\pm$1.2 & 93.1$\pm$1.1 & 94.3$\pm$1.0 & 94.9$\pm$0.7 & 59.5$\pm$4.2 & 86.0$\pm$0.6 & \textbf{91.9$\pm$1.2} & 93.6$\pm$1.2 & \textbf{94.7$\pm$0.8} \\
		I+R & & \textbf{69.1$\pm$1.3} & 87.1$\pm$0.6 & \textbf{93.5$\pm$0.7} & \textbf{94.6$\pm$0.6} & 95.0$\pm$0.7 & \textbf{62.0$\pm$1.1} & \textbf{87.2$\pm$1.1} & \textbf{91.9$\pm$0.7} & \textbf{93.9$\pm$0.9} & 94.5$\pm$1.0 \\
		F+R & & 14.6$\pm$0.9 & 44.1$\pm$1.1 & 76.1$\pm$0.7 & 89.5$\pm$0.3 & 93.8$\pm$0.3 & 16.5$\pm$0.7 & 43.7$\pm$1.6 & 72.4$\pm$0.7 & 85.2$\pm$1.5 & 91.1$\pm$0.6 \\ \midrule
		I+F+R & 0.090 & 68.5$\pm$2.0 & \textbf{88.0$\pm$0.8} & 92.8$\pm$1.2 & 94.2$\pm$1.0 & \textbf{95.1$\pm$1.0} & 59.9$\pm$3.3 & 86.3$\pm$1.9 & 91.8$\pm$1.1 & 93.5$\pm$0.6 & 94.4$\pm$0.7 \\
		\midrule[0.3pt]\bottomrule[1pt]
	\end{tabular}
\end{table*}

\section{Experiments and Results}
\label{sec:results}

In order to establish a baseline for noise-robust HA voice control, this section presents KWS accuracy results from using the different HA microphones as well as their combinations. KWS accuracy (which is the standard metric for the GSCD \cite{KWSOverview}) is reported along with 95\% confidence intervals from the Student's $t$-distribution \cite{Blachman87}. Given an experiment, to obtain its confidence interval, 5 distinct \texttt{BC-ResNet} models were trained by using different random parameter initialization.

Fig. \ref{fig:single} shows KWS accuracy (averaged across both SNRs and 5 trained models) as a function of the number of model parameters ---which is determined by the value of $\tau$ (see Subsec. \ref{ssec:model})--- when evaluating the use of the different HA microphones. The resulting curves are also broken down by whether noises were \textbf{seen} (babble, music, and SSN) or \textbf{unseen} (interfering speaker and TV) during training. As expected, KWS performance is better on the seen-noise test partition than on the unseen-noise one. Furthermore, while KWS accuracy resulting from using the front microphone is comparable to that from using the rear one, performance is clearly superior when employing the IEC microphone presumably due to the resilience of BCS to ambient noise.

Fig. \ref{fig:combo} depicts KWS accuracy curves from the combination of the different HA microphones. IEC curves are also displayed as a reference. We can see from this figure that combining BCS from the IEC microphone with regular air-conducted speech from the BTE microphones boosts KWS performance. We conjecture that this is due to the noisy higher frequency information from the BTE microphone signals complementing noise-robust, bandwidth-limited BCS. Moreover, we can also observe from Fig. \ref{fig:combo} that I+F, I+R and I+F+R yield a comparable accuracy, which probably is due to the employed front- and rear-microphone spectral representations containing relatively similar information (see Fig. \ref{fig:specs}) given that the BTE microphones are close each other. This hypothesis is partially endorsed by F+R providing relatively small accuracy gains with respect to the front and rear microphones in isolation (see Fig. \ref{fig:combo} versus Fig. \ref{fig:single}).

Finally, note from Fig. \ref{fig:combo} that KWS performance increases with the number of model parameters and starts to stagnate from 50k parameters. This makes \texttt{BC-ResNet-3} (which has 54,168, 55,368 and 56,568 parameters for 1-, 2- and 3-microphone inputs, respectively) a good candidate to be deployed in HAs taking into account that they are low-resource devices. Table \ref{tab:results} shows detailed KWS accuracy results (broken down by SNR and seen/unseen noises) and real-time factors (RTFs)\footnote{RTFs, defined as the ratio between the time taken to process an input and the input duration \cite{Pratap20}, were measured on an Intel Xeon Silver 4214R CPU with a clock frequency of 2.4 GHz.} obtained by \texttt{BC-ResNet-3} employing the previously tested microphones and combinations of them. From this table, we can clearly assess how the combination of noise-robust BCS with the broader-bandwidth BTE microphone signals yields good voice control performance at extremely challenging SNRs.

\section{Conclusion}
\label{sec:conclusion}

In this work, we have aimed at establishing a baseline for noise-robust HA voice control. This has been accomplished through \emph{1)} the creation of a publicly available\footnotemark[2] HA noisy dataset comprising both regular air- and bone-conducted own voice, and \emph{2)} the evaluation of light, state-of-the-art KWS models for HA voice control using such a dataset. Experimental results have demonstrated that it is possible to achieve robust voice control by exploiting BCS from an IEC microphone as well as voice control performance can be boosted by adding complementary noisy higher frequency information from BTE microphones.

Future work will investigate the collection of and evaluation on real noisy own voice, since the available own-voice data, which were used for OVTF estimation, are clean only.

\bibliographystyle{IEEEbib}
\bibliography{strings,refs}

\end{document}